\documentclass[lettersize,journal]{IEEEtran}
\usepackage{cite}
\usepackage{cite,balance}
\usepackage{cancel}
\usepackage{multicol}
\usepackage{amsmath,amssymb,amsfonts,nicematrix}
\usepackage{bbm}
\usepackage{xcolor}
\usepackage{algorithm}
\usepackage{algpseudocode}
\def\BibTeX{{\rm B\kern-.05em{\sc i\kern-.025em b}\kern-.08em
    T\kern-.1667em\lower.7ex\hbox{E}\kern-.125emX}}
\usepackage{nicematrix,tikz}
\usepackage{graphicx}
\usepackage{textcomp}
\usepackage{array,booktabs,makecell}
\usepackage[font=footnotesize]{caption}
\usepackage{float}
\usepackage{tikz}
\usepackage{cite,balance}
\usetikzlibrary{shapes.geometric, arrows}
\usepackage{pgfplots}
\usepackage{caption}
\usepackage{mathtools}
\usepackage{subcaption}
\usepackage{comment}
\usepackage{enumitem}
\usepackage{stackengine}
\newcommand\numberthis{\addtocounter{equation}{1}\tag{\theequation}}
\usepackage{xcolor}
\definecolor{OliveGreen}{rgb}{0,0.6,0}
\def\BibTeX{{\rm B\keri-.05em{\sc i\keri-.025em b}\keri-.08em
    T\keri-.1667em\lower.7ex\hbox{E}\keri-.125emX}}
\usepackage{multirow}

\usepackage{amsthm}

\newtheorem{definition}{Definition}
\newtheorem{theorem}{Theorem}

\newtheorem{lemma}{Lemma}

\allowdisplaybreaks
\usepackage{mathtools}

\def\delequal{\mathrel{\ensurestackMath{\stackon[1pt]{=}{\scriptstyle\Delta}}}}

\setlist[enumerate]{wide=0pt, leftmargin=15pt, labelwidth=15pt, align=left}
\usepackage{tabstackengine}
\setstackEOL{\cr}
\definecolor{darkgray}{rgb}{0.66, 0.66, 0.66}
\usetikzlibrary{spy,calc}
\usepgfplotslibrary{units}
\usetikzlibrary{spy,backgrounds}
\usepackage{pgfplotstable}
\pgfplotstableread{
0.0    1.0
0.0   -0.5
}\datatable
\makeatletter
\tikzset{spy on other/.code={
  \pgfutil@g@addto@macro\tikz@lib@spy@collection{%
    \setbox\tikz@lib@spybox=\hbox{\pgfpicture#1\endpgfpicture}}}}
\makeatother
\usepackage{mathtools}

\DeclarePairedDelimiter\floor{\lfloor}{\rfloor}
\definecolor{capri}{rgb}{0.0, 0.75, 1.0}
\definecolor{bleudefrance}{rgb}{0.19, 0.55, 0.91}
\definecolor{bronze}{rgb}{0.8, 0.5, 0.2}
\definecolor{burntorange}{rgb}{0.8, 0.33, 0.0}
\definecolor{cardinal}{rgb}{0.77, 0.12, 0.23}
\definecolor{bazaar}{rgb}{0.6, 0.47, 0.48}
\definecolor{darklavender}{rgb}{0.45, 0.31, 0.59}
\definecolor{dollarbill}{rgb}{0.52, 0.73, 0.4}
\definecolor{dodgerblue}{rgb}{0.12, 0.56, 1.0}
    
\begin{document}

\title{Coded Cooperative Networks for Semi-Decentralized Federated Learning}

\author{Shudi Weng,~\IEEEmembership{Graduate Student Member,~IEEE,}
Ming Xiao,~\IEEEmembership{Senior Member,~IEEE,}\\
Chao Ren,~\IEEEmembership{Member,~IEEE,}
and Mikael Skoglund,~\IEEEmembership{Fellow,~IEEE}\vspace{-2em}
\thanks{
This work was partly supported by the SUCCESS project (FUS21-0026), funded by the Swedish Foundation for Strategic Research.

Shudi Weng, Ming Xiao, Chao Ren, and Mikael Skoglund are with the School of Electrical Engineering and Computer Science (EECS), KTH Royal Institute of Technology, 11428 Stockholm, Sweden, Email: \{shudiw, mingx, chaor, skoglund\}@kth.se.}
}
\markboth{Journal of \LaTeX\ Class Files,~Vol.~14, No.~8, August~2021}%
{Shell \MakeLowercase{\textit{et al.}}: A Sample Article Using IEEEtran.cls for IEEE Journals}


\maketitle

\begin{abstract}
To enhance straggler resilience in federated learning (FL) systems, a semi-decentralized approach has been recently proposed, enabling collaboration between clients. Unlike the existing semi-decentralized schemes, which adaptively adjust the collaboration weight according to the network topology, this letter proposes a deterministic coded network that leverages wireless diversity for semi-decentralized FL without requiring prior information about the entire network. Furthermore, the theoretical analyses of the outage and the convergence rate of the proposed scheme are provided. Finally, the superiority of our proposed method over benchmark methods is demonstrated through comprehensive simulations. 
\end{abstract}

\begin{IEEEkeywords}
Semi-decentralized federated learning, Wireless network, Diversity network code, Communication stragglers
\vspace{-1mm}
\end{IEEEkeywords}

\section{Introduction}
\IEEEPARstart{F}{ederated} learning (FL) is a promising distributed edge learning paradigm that leverages the local computational capabilities of edge devices to exploit its local datasets, 
by iteratively optimizing a common objective function and collaborating with the central parameter server (PS) to learn a global model\cite{wen2023survey}. In FL, clients collect and store the training datasets locally, thereby greatly decreasing the volume of data transmitted during the training process and preserving data privacy by avoiding raw dataset sharing. 
However, the resulting heterogeneous data distribution across clients can potentially hinder the convergence of FL or even lead to strict sub-optimality of FL algorithms, if not properly managed, as the non-i.i.d. (independently and identically distributed) data stored locally on a subset of clients does not precisely reflect the overall population distribution. Therefore, the cases involving partial client participation must be carefully designed\cite{ye2023heterogeneous}. 

Most existing works on FL with data dissimilarity consider perfect links by assuming that the wireless imperfections have been handled by communication protocols \cite{chen2022federated, bouzinis2023wireless}. However, in realistic FL scenarios, some clients may fail to connect with PS due to physical factors like fading, shadowing, and resource constraints. Such clients, known as \textit{communication stragglers}, are unable to regularly update their local models, leading to partial client participation and degraded FL performance. Therefore, a separate design between the communication system and learning process can be strictly sub-optimal\cite{yemini2023robust}.

\vspace{0.3em}
\subsubsection{Related works} 
There are multiple strategies to improve the straggler resilience in FL. The diverse physical environments may induce different possibilities of clients being stragglers. 
The impact of this different connectivity on FL convergence is investigated in \cite{wang2021quantized}, which proposes an adaptive resource allocation method to eliminate the objective inconsistency induced by stragglers and thus shrink the optimality gap. 
Alternatively, \cite{vu2021straggler} designs client sampling strategies to reduce the probability of sampling a straggler in FL over wireless networks with intermittent client availability and enhance FL convergence performance,   
\begin{figure}
    \centering
    \includegraphics[width=0.7\linewidth]{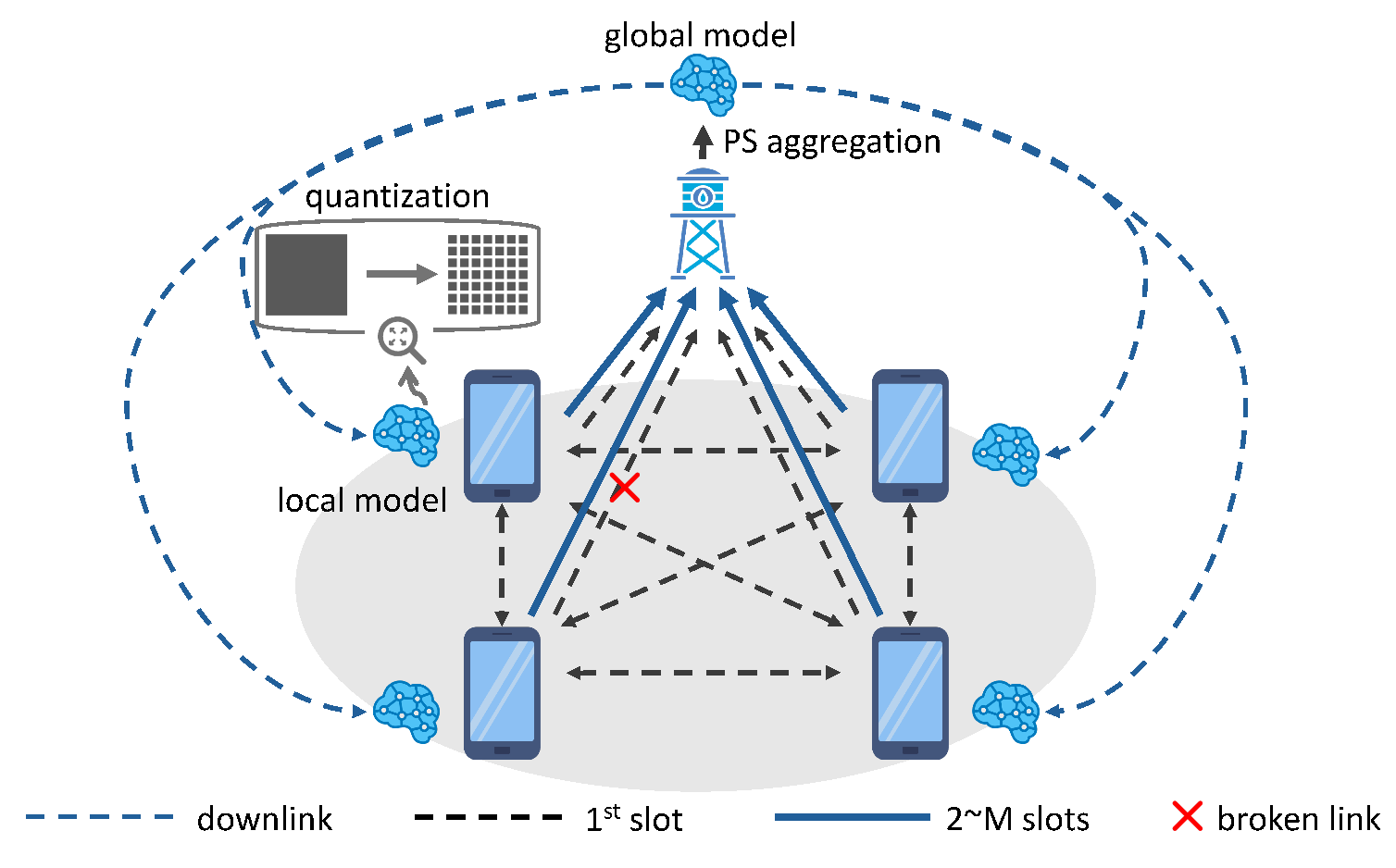}
    \vspace{0mm}
    \caption{Illustration of the proposed scheme within the semi-decentralized FL system over the intermittent links in $M$ slots with two communication stages.}
    \label{fig:senario}
    \vspace{-3mm}
\end{figure}
and \cite{yemini2022semi} proposes to alter the network topology to address stragglers by enabling communication between clients, rather than relying on sole communication between clients and PS, 
referred to as \textit{semi-decentralized FL}, or \textit{collaborative FL}. 
However, the unbiased estimation of the true global model at PS in the aforementioned methods\cite{wang2021quantized,vu2021straggler,yemini2022semi} is contingent upon precise prior information, such as entire network connectivity, client availability, and time synchronization, 
which markedly amplifies implementation complexity in real-world scenarios. 

\vspace{0.3em}
\subsubsection{Our contributions}
To overcome the limitations of the existing methods, this letter proposes a novel deterministic cooperative network that is straggler-resilient and does not necessitate prior information about the network, and enables the retrieval of individual local models from a subset of clients. Our contributions are summarised as follows.
\begin{itemize}
    \item We propose a deterministic coded cooperative networking scheme for semi-decentralized FL based on the maximum distance separable (MDS) code exploiting wireless diversity. To the best of our knowledge, this is the first work to exploit coded diversity to mitigate stragglers in FL.
    \item We conduct rigorous theoretical analyses of the proposed scheme. Specifically, we provide the outage analysis of PS not being able to see local model updates from each client through the intermittent network and the convergence rate analysis of the proposed scheme.
    \item We verify the effectiveness of the proposed scheme by simulations and comparisons with benchmark methods. 
\end{itemize}

\section{System Model for Semi-decentralized FL \\over Wireless Network}


The FL system typically consists of a central PS and multiple clients. Let $\mathcal{L}(\boldsymbol{\theta},\xi)$ be the loss evaluated for a model $\boldsymbol{\theta}$ at a data sample $\xi$. Denote the local dataset on client $m\in [M]$ as $\mathcal{D}_m$, and its local objective function as $F_m: \mathbb{R}^d\times \mathcal{D}_m\rightarrow \mathbb{R}$, where $F_m(\boldsymbol{\theta},\mathcal{D}_m)=\frac{1}{\lvert \mathcal{D}_m \rvert}\sum_{\xi\in \mathcal{D}_m}\mathcal{L}(\boldsymbol{\theta},\xi)$. 
The objective of the entire FL system is to solve the following empirical risk minimization (ERM) problem collaboratively:
\begin{align*}
    \min_{\boldsymbol{\theta}\in \mathbb{R}^d}\left\{F(\boldsymbol{\theta})\delequal \frac{1}{M}\sum_{m=1}^{M}F_m(\boldsymbol{\theta}, \mathcal{D}_m)\right\},
    \numberthis
\end{align*}
where $F(\cdot)$ is the global objective function. We assume identical importance for all clients, i.e., $\lvert \mathcal{D}_m \rvert=\lvert \mathcal{D}_k \rvert$, $\forall k, m\in [M] $.

\subsection{Local Training at Clients}\label{sec: communication model}
Define the true gradient of the local objective function as the gradient $\nabla F_m(\boldsymbol{\theta}, \mathcal{D}_m)$ over the entire local dataset. 

At the beginning of the $r$-th round, clients initialize with the latest global model received from PS, i.e., $\boldsymbol{\theta}_{m,r}^0=\boldsymbol{\theta}_{r-1}$. Subsequently, each client performs $I$-step local SGD  and updates the local model at the $i$-th iteration as
\begin{align*}
\boldsymbol{\theta}_{m,r}^i\leftarrow\boldsymbol{\theta}_{m,r}^{i-1}-\eta\nabla F_m(\boldsymbol{\theta}_{m,r}^{i-1},\boldsymbol{\xi}_{m,r}^{i}),\;\;\; i\in[I],
    \numberthis
    \label{eq:Local_update}
\end{align*}
where $\boldsymbol{\xi}_{m,r}^{i}$ is the corresponding training data patch randomly extracted from the local dataset $\mathcal{D}_m$ at the $i$-th iteration of the $r$-th training round, and $\nabla F_m(\boldsymbol{\theta}_{m,r}^{i-1},\boldsymbol{\xi}_{m,r}^{i})$ is the stochastic gradient, i.e., the stochastic estimation of the true gradient. 

\subsection{Transmission over Wireless Network}\label{sec: communication model}
The scenario is depicted in Fig. \ref{fig:senario}. Clients collect edge data, perform local training, generate network codewords, and manage communications. PS manages communication with clients, decodes both the individual messages and network codewords sent by clients and computes the global model. 
\subsubsection{Network Model}
The semi-decentralized network involves two stages: \textit{device-to-device (D2D) communication} and \textit{device-to-PS (D2P) communication}.  
The clients convey their information to PS via wireless medium with the help of their neighbors. Any link may suffer disruption and fail to update with PS. The intermittent D2D network can be captured by the random binary matrix $\boldsymbol{\mathcal{T}}(r)\in\{0, 1\}^{M\times M}$, whose $(m, k)$-th entry $\tau_{mk}(r)\sim \mathrm{Bernoulli}(1-q_{mk})$, where $q_{mk}$ is the outage probability of the link from client $m$ to client $k$ and $q_{mm}=0$ for every $m\in [M]$ since there is no transmission. The D2P network can be captured by the binary random vector $\boldsymbol{\tau}(r)\in\{0, 1\}^{M\times 1}$, whose $m$-th entry $\tau_{m}(r)\sim \mathrm{Bernoulli}(1-q_{m})$, where $q_{m}$ is the outage probability of the link from client $m$ to PS. Notably, the discussion of scheduling and interference in multi-access channels is beyond the scope of this paper.




\subsubsection{Quantized Transmission and Outage}
Next, we briefly describe the transmission process and the outage model for an individual link. Before transmitting to other devices and PS, device $m$ needs to quantize $\Delta\boldsymbol{\theta}_{m,r}^{I}\in \mathbb{R}^d$ such that a finite number of symbols can represent the source. The most popular compression technique employed in learning systems is stochastic quantization (SQ) \cite{wang2021quantized,amiri2020federated}, whose characteristic function is given in (\ref{eq:SQ}). For any given number $\Delta\theta\in \Delta\boldsymbol{\theta}_{m,r}^{I}$,
\begin{align*}
    \mathcal{Q}(\Delta\theta)=\begin{cases}
\floor{\Delta\theta},\;\;\;\;\;\;\;\;\;\;\;\mathrm{w.p.}\;\;\frac{\floor{\Delta\theta}+\kappa-\Delta\theta}{\kappa}\\    \floor{\Delta\theta}+\kappa,\;\;\;\;\;\mathrm{w.p.}\;\;\frac{\floor{\Delta\theta}-\Delta\theta}{\kappa},
    \end{cases}
    \numberthis
    \label{eq:SQ}
\end{align*}
where $\floor{\Delta\theta}$ is the largest multiple of $\kappa$ such that $\floor{\Delta\theta}\leq \Delta\theta$, and $\kappa$ is the interval length of uniformly distributed knobs \cite{amiri2020federated,wang2021quantized}. W.L.O.G., we assume all clients equipped with the same stochastic quantizer and encoder $\mathcal{E}: \mathbb{R}^d\rightarrow \mathbb{F}_p^k$ that maps $\mathcal{Q}(\Delta\boldsymbol{\theta}_{m,r}^{I})$ into finite-field massage $U_{m,r}$ according to the default systematic Gaussian codebook, i.e., 
\begin{align*}
    U_{m,r}=\mathcal{E}\left(\mathcal{Q}(\Delta\boldsymbol{\theta}_{m,r}^{I})\right).
        \numberthis
    \label{eq:encoder}
\end{align*}

For simplicity of theoretical analysis, all wireless links are assumed to be independent and identically distributed (i.i.d.) block fading channels. Assume all clients transmit their message/network codewords at rate $R$ under signal-to-noise ratio (SNR) $\mathrm{SNR}$ through orthogonal access. When a client hears from another client, it performs maximum-ratio combining (MRC) first and then decodes the corresponding codeword. Specifically, let $h_{mk}$ denote the fading channel gain from client $m$ to client $k$, client $k$ cannot decode $U_{m,r}$ correctly when the channel capacity is less than the transmission rate, i.e., when $C=\frac{1}{2}\log(1+\lvert h_{mk}\rvert^2\mathrm{SNR})<R$. Or equivalently, when $\lvert h_{mk}\rvert^2<g$, where $g=\frac{2^{2R}-1}{\mathrm{SNR}}$.
Assume Rayleigh fading, i.e., $ h_{mk}\sim \mathcal{CN}(0,\sigma^2)$, where $\mathcal{CN}(0,\sigma^2)$ is zero-mean complex Gaussian distribution with variance $\sigma^2$. Then 
the outage probability $q_{mk}$ per transmission is given by $q_{mk}=1-e^{-g/2\sigma^2}$. For ease of reading, let $P_e$ represent $1-e^{-g/2\sigma^2}$ in the following. If the receiver can successfully recover $U_{m,r}$, then the decoder $\mathcal{E}^{-1}: \mathbb{F}_p^k\rightarrow \mathbb{R}^d$ allows the receiver to map the quantized learning model back, as
\begin{align*}
    \mathcal{Q}(\Delta\boldsymbol{\theta}_{m,r}^{I})=\mathcal{E}^{-1}\left(U_{m,r}\right).
    \numberthis
\end{align*}

\subsection{Aggregation at PS}
Ideally, PS aggregation aims at computing $\frac{1}{M}\sum_{m=1}^M\boldsymbol{\theta}_{m,r}^I$. However, full client participation is unrealistic in large-scale FL. If we express the aggregation resulting from any algorithm by the function $\mathcal{S}: (\boldsymbol{\theta}_{1,r}^I, \cdots, \boldsymbol{\theta}_{M,r}^I)\rightarrow \boldsymbol{\theta}_{r}$, the alternative goal of PS aggregation with partial client participation is to estimate the unbiased global model statistically, i.e.,
\begin{align*}
\mathbb{E}_{\boldsymbol{\mathcal{T}}(r), \boldsymbol{\tau}(r)}\left[\mathcal{S}(\boldsymbol{\theta}_{1,r}^I, \cdots, \boldsymbol{\theta}_{M,r}^I)\right]=\frac{1}{M}\sum_{m=1}^M\boldsymbol{\theta}_{m,r}^I,
\numberthis
\end{align*}
where $\mathbb{E}[\cdot]$ is taken over stochasticity of the intermittent network captured by $\boldsymbol{\mathcal{T}}(r)$ and $\boldsymbol{\tau}(r)$. 

\section{The Proposed Method: Coded Cooperative Network for Semi-decentralized FL}
\vspace{-0.5em}
In this section, we describe the proposed coded cooperative networking scheme \footnote{In cooperative communication, clients are typically assumed to trust each other. The discussion of privacy is out of the scope of this letter.} in semi-decentralized FL. The employed network coding scheme, termed \textit{diversity network code (DNC)}, is first proposed in \cite{5595117} to enhance the robustness of the cooperative communication. 
\subsection{System Description}
Assume all clients and PS 
can decode each other's message. Let $T$ be the total number of communication rounds. Here, we describe the proposed scheme at the $r$-th round with 4 stages. 
\subsubsection{Broadcasting} In the beginning, PS broadcasts the latest global model $\boldsymbol{\theta}_{r-1}$ to all clients. For simplicity, the downlink channels are assumed to be error-free.
\subsubsection{Local training} Each client initializes its local model by setting $\boldsymbol{\theta}_{m,r}^{0}=\boldsymbol{\theta}_{r-1}$, and performs $I$-step iterative local training as in  (\ref{eq:Local_update}). 
\subsubsection{Transmission} 
After completing the local training, the local model update $\Delta\boldsymbol{\theta}_{m,r}^{I}\in \mathbb{R}^d$ on each client is quantized by stochastic quantizer as in (\ref{eq:SQ}). 
Subsequently, encoder $\mathcal{E}$ maps $\mathcal{Q}\left(\Delta\boldsymbol{\theta}_{m,r}^{I}\right)$ to finite-field message $U_{m,r}$ as in (\ref{eq:encoder}) according to the provided systematic codebook. Now clients are ready to 
perform the following two stages of communication.

\vspace{1mm}
\textbf{$\mathbf{1}$st time slot:}  
    Each client transmits its message via a distinct orthogonal frequency slot. Due to the broadcasting nature of the wireless medium, both the PS and other clients can potentially hear from client $m$ and may decode $U_{m,r}$ depending on connectivity resulting from the channel condition. 

\vspace{1mm}
\textbf{$\mathbf{2\sim M}$ time slots:}  
    After attempting to decode all messages heard from other clients, client $m\in[M]$ generates $M-1$ network codewords for transmissions in the next $2\sim M$ time slots over the $M$ orthogonal frequency slots allocated for $M$ client. For ease of reading, let us temporarily assume that the network encoding matrix $\boldsymbol{A}$ in the form of (\ref{eq:DNC code}) exists, allowing us to clearly describe our proposed method. This fact will be justified later. Let $\boldsymbol{A}_m$ denote the encoding block of client $m$, if client $m$ can decode all messages from other clients, then it generates the network codewords for $2\sim M$ slots as
\begin{align*}
    \boldsymbol{C}_{m,r}= \boldsymbol{U}_r \boldsymbol{A}_m,
    \numberthis
\end{align*}
where $\boldsymbol{U}_r=[U_{1,r}, \cdots, U_{M,r}]$ is the collection of individual messages, and $\boldsymbol{C}_{m,r}$ contains the generated $M-1$ network codewords.
However, if client $m$ fails to decode some $U_{z,r}$ from client $z$, 
it sets the $z$-th row in $\boldsymbol{A}_m$ by $0$s
before generating network codewords. The resulting encoding matrix of client $m$ is denoted by $\tilde{\boldsymbol{A}}_{m,r}$. Let $\boldsymbol{\mathcal{T}}(r)$ and $\boldsymbol{\tau}(r)$ denote the binary connectivity matrix \textit{between devices} and \textit{between devices and PS} respectively in the 1st slot.  Additionally, let $\boldsymbol{\tau}^{(m)}(r)$ denote the $m$-th column of $\boldsymbol{\mathcal{T}}(r)$, that is, $\boldsymbol{\tau}^{(m)}(r)$ represent the binary connectivity from other clients to client $m$. 
Then, the above process can be expressed as
\begin{align*}
    &\tilde{\boldsymbol{A}}_{m,r}= \boldsymbol{A}_m \odot \left( \mathbf{1}_{M-1}^\top \otimes \boldsymbol{\tau}^{(m)}(r) \right),
    \numberthis
    \label{eq: encoding_cleint_m}
\end{align*}
where $\odot$ is the column-wise Khatri-Rao product, $\otimes$ is Kronecker product, $\mathbf{1}_{M-1}$ is all-one vector of size $M-1$. Then the actually generated network codewords sent from client $m$ in $2\sim M$ slots can be expressed as
\begin{align*}
    \tilde{\boldsymbol{C}}_{m,r}= \boldsymbol{U}_r \tilde{\boldsymbol{A}}_{m,r}.
    \numberthis
\end{align*}


Note that we assume the resulting total of $M(M-1)$ network codewords are transmitted over orthogonal channels, with $M$ orthogonal frequency slots allocated for $M$ clients, and $M-1$ time slots assigned to each client. In practice, however, these orthogonal slots can be flexibly assigned across time and frequency domains to balance various practical communication demands, such as bandwidth and delay\footnote{The number of decoding operations is approximately proportional to $M^2$, however, with proper orthogonal frequency slot allocation and parallel decoding structure at each device, the minimal decoding time can be 2 slots.}.

Let $\boldsymbol{\tau}_m(r)$ represent the binary connectivity vector from client $m$ to PS in $2\sim M$ slots. Then the received codewords from client $m$ at PS in $2\sim M$ slots are 
\begin{align*}
     \bar{\boldsymbol{C}}_{m,r}= \tilde{\boldsymbol{C}}_{m,r} \odot \boldsymbol{\tau}_m^\top(r).
    \numberthis
    \label{eq:final_words_at_PS}
\end{align*}
As a result, the actual encoding matrix $\hat{\boldsymbol{A}}_{m,r}$ of the finally received network codewords from client $m$ at PS is
\begin{align*}
    &\hat{\boldsymbol{A}}_{m,r}=\tilde{\boldsymbol{A}}_{m,r} \odot \boldsymbol{\tau}_m^\top(r),\\
    &\hspace{6mm}= \boldsymbol{A}_m \odot \left( \mathbf{1}_{M-1}^\top \otimes \boldsymbol{\tau}^{(m)}(r) \right)\odot 
    \boldsymbol{\tau}_m^\top(r),
    \numberthis
\end{align*} 
and the entire encoding matrix $\hat{\boldsymbol{A}}_r$ of all received network codewords at PS is
\begin{align*}
    &\hat{\boldsymbol{A}}_r=\left[ \boldsymbol{I}_M\cdot \mathrm{Diag}\left\{\boldsymbol{\tau}(r)\right\}, \hat{\boldsymbol{A}}_{1,r}, \cdots, \hat{\boldsymbol{A}}_{M,r} \right].
    \numberthis
\end{align*}

\subsubsection{PS Decoding and Aggregation}
Let $\bar{\boldsymbol{A}}_r$ denote $\hat{\boldsymbol{A}}_r$ but excluding all-zero rows and columns, i.e., 
\begin{align*}
    &\mathcal{W}_r=\{w\in[M]: \boldsymbol{\alpha}_{w}\neq\mathbf{0}^\top_{M^2}\},
    \numberthis\\
    &\mathcal{V}_r=\{v\in[M^2]: \boldsymbol{\beta}_{v}\neq\mathbf{0}_{M}\},
    \numberthis\\
    &\bar{\boldsymbol{A}}_r=\hat{\boldsymbol{A}}_r(\mathcal{W}_r,\mathcal{V}_r),
    \numberthis
    \label{eq:Bar_A}
\end{align*}
where $\boldsymbol{\alpha}_{w}$ and $\boldsymbol{\beta}_{v}$ denote the $w$-th row and the $v$-th column in $\hat{\boldsymbol{A}}_r$, respectively. 

Let $\Bar{\boldsymbol{U}}_r=\boldsymbol{U}_r(\mathcal{W}_r)$ be the collection of the involved individual messages in the network codewords that arrived at PS, denoted by $\bar{\boldsymbol{C}}_r$, according to (\ref{eq: encoding_cleint_m})$\sim$(\ref{eq:Bar_A}) we have 
\begin{align*}
    \bar{\boldsymbol{C}}_r=\Bar{\boldsymbol{U}}_r \bar{\boldsymbol{A}}_r.
    \numberthis
\end{align*}
 If $\lvert \mathcal{W}_r\rvert \geq \lvert \mathcal{V}_r \rvert$, $\bar{\boldsymbol{A}}_r$ is overdetermined, PS can decode $\Bar{\boldsymbol{U}}_r$. That is, PS can decode the messages from clients in set $\mathcal{W}_r$. If $\lvert \mathcal{W}_r\rvert < \lvert \mathcal{V}_r \rvert$ or $\mathcal{W}_r = \emptyset$, repeat communication until $\lvert \mathcal{W}_r\rvert \geq \lvert \mathcal{V}_r \rvert$. Thus, PS can decode messages from client $m\in\mathcal{W}_r$.

Then PS aggregates the received local model updates as 
\begin{align*}
    \Delta\boldsymbol{\theta}_{r}\leftarrow \sum_{m\in \mathcal{W}_r} \frac{1}{\lvert \mathcal{W}_r\rvert} \Delta\boldsymbol{\theta}_{m,r}^I.
    \numberthis
    \label{eq:global_update_partial}
\end{align*}
This update rule corresponds to scheme II in \cite{li2019convergence}, the unbiasedness and advantage of (\ref{eq:global_update_partial}) is discussed in Lemma 1. The $r$-th round global model is updated as $\boldsymbol{\theta}_{r}\leftarrow \boldsymbol{\theta}_{r-1}+\Delta\boldsymbol{\theta}_{r}$.


\begin{figure*}[!htb]
  \centering
\begin{equation*}
    \boldsymbol{A}=\begin{bNiceMatrix}
        1 & 0& \cdots & 0 &\alpha_{1,1} & \cdots & \alpha_{1,M-1} & \cdots & \alpha_{1,(M-1)(M-1)+1} & \cdots & \alpha_{1,M(M-1)}\\ 
        0& 1 & \ddots & \vdots &\alpha_{2,1} & \dots & \alpha_{2,M-1} & \cdots& \alpha_{2,(M-1)(M-1)+1} & \cdots & \alpha_{2,M(M-1)}\\
        \vdots &\ddots & \ddots & 0 &\vdots &  & \vdots & & \vdots & & \vdots\\
        0 & \cdots &0 & 1 &\alpha_{M,1} & \cdots & \alpha_{M,M-1} & \cdots & \alpha_{1,(M-1)(M-1)+1} & \cdots & \alpha_{M,M(M-1)}\\
    \CodeAfter
    \OverBrace[shorten,yshift=5pt]{1-5}{1-7}{\boldsymbol{A}_1}
    \OverBrace[shorten,yshift=5pt]{1-9}{1-11}{\boldsymbol{A}_M}
    \end{bNiceMatrix}
    \numberthis
    \label{eq:DNC code}
\end{equation*}
  \vspace{-1.5em}
\end{figure*}
\subsection{Network Code Design}
 If client $m$ transmits message $U_{m,r}$ to PS, due to the broadcasting nature of the wireless medium, other clients may also receive $U_{m,r}$. Let $\hat{\mathcal{N}}_m$ denote the set of clients that can decode and relay $U_{m,r}$, then the network codewords transmitted from ${\mathcal{N}}_m=\hat{\mathcal{N}}_m \cup \{ \mathrm{client}\; m \}$ will involve $U_{m,r}$.
\begin{definition}
DNC is any deterministic network code used as described in Section III-A in \cite{5595117} such that PS can recover $U_{m,r}$ if it can decode any $\lvert \mathcal{N}_m\rvert$ different network codewords out of total $M \lvert \mathcal{N}_m\rvert$ codewords from clients in ${\mathcal{N}}_m$. 
\end{definition}

By Prop. 1 in \cite{5595117}, DNC for semi-decentralized FL systems with $M$ clients exists. Furthermore, Section III-D in \cite{5595117} provides a simplified DNC construction in the form of (\ref{eq:DNC code})  without loss of performance based on MDS code construction. The finite field of size $M^2-1 \choose M-1$ is sufficient for the construction. It can be verified every sub-matrix of $M$ columns in (\ref{eq:DNC code}) is of full rank since any $M$ columns have rank $M$.



\section{Performance Analysis}
\subsection{Outage Analysis}
The chances of clients conveying their local model updates to PS are equal through a symmetric intermittent network with i.i.d. links. This can be viewed as \textit{uniform sampling} in \cite{li2019convergence} but induced by network connectivity.
\cite{5595117} provided a thorough outage analysis of DNC. For ease of reference, we briefly summarize it and adapt it to our case. The outage probability of PS not being able to see local model updates from client $m$ is {dominated} by $P_e^{2M-1}$, which corresponds to the scenario where no other clients can decode and relay $U_{m,r}$. The outage probability is monotonously decreasing with the number of clients that can decode and relay $U_{m,r}$. The {dominance} for the probability of outages of individual local model updates is sufficient to gain insight into the convergence rate.

\subsection{Convergence Analysis}
We conduct a non-convex convergence analysis for the proposed scheme under the following assumptions\cite{9515709}, \cite{wang2020tackling}. 
\begin{itemize}
    \item[A.1] (Smoothness) 
    Each local objective function is bounded by $F_m(x)\geq F^\star$ and is differentiable. Its gradient $\nabla F_m(x)$ is L-smooth, i.e., $\lVert \nabla F_m(x)-\nabla F_m(y) \rVert\leq L\lVert x-y \rVert$, $\forall i \in [M]$.
    \item[A.2] (Unbiased gradient and bounded data variance) 
    The local stochastic gradient is an unbiased estimation, i.e., $\mathbb{E}_\xi[\nabla F_m(x,\xi)]=\nabla F_m(x)$, and has bounded data variance $\mathbb{E}_\xi[\lVert \nabla F_m(x,\xi)-\nabla F_m(x)\rVert^2]\leq \sigma^2 $, $\forall i \in [M]$. 
    \item[A.3] (Bounded data dissimilarity)
    The dissimilarity between $\nabla F_m(x)$ and $\nabla F(x)$ is bounded, i.e., $\mathbb{E}[\lVert \nabla F_m(x)-\nabla F(x)\rVert^2]\leq D_m^2$, $\forall i \in [M]$. 
\end{itemize}

Next, we present two key lemmas to acquire Theorem 1.
\begin{lemma}
Given that PS is equally likely to see each local model update from any client with $1-q$, and the aggregation of these updates to recover the global model is statistically unbiased in terms of the expected value, that is, 
\begin{align*}   
\mathbb{E}_{\mathcal{W}_r}\left[\sum_{m\in \mathcal{W}_r} \frac{1}{\lvert \mathcal{W}_r\rvert}  \Delta\boldsymbol{\theta}_{m,r}^I\Bigg \vert \mathcal{W}_r\neq \emptyset\right]=\sum_{m=1}^M \frac{1}{M}\Delta\boldsymbol{\theta}_{m,r}^I.
\numberthis
\end{align*}
Besides, it can be proved that   
\begin{align*}   
\mathbb{E}_{\mathcal{W}_r}\left[\sum_{m\in \mathcal{W}_r} \frac{1}{\lvert \mathcal{W}_r\rvert^2} \Delta\boldsymbol{\theta}_{m,r}^I\Bigg \vert \mathcal{W}_r\neq \emptyset\right]\triangleq\sum_{m=1}^M \Bar{\alpha}_m \Delta\boldsymbol{\theta}_{m,r}^I,
\numberthis
\end{align*}
where $\Bar{\alpha}_m=\frac{1}{M\Bar{K}}$ with $\frac{1}{\Bar{K}}=\sum_{l=1}^M\frac{\frac{1}{l}\mathbb{C}_M^l(1-q)^lq^{M-l}}{1-q^M}\leq \frac{2}{(M+1)(1-q)(1-q^M)}\approx \frac{2}{(M+1)(1-P_e^{2M-1})(1-P_e^{M(2M-1)})} \triangleq \frac{1}{K^{\star}}$ since $q\approx P_e^{2M-1}$. 
\begin{proof}
    The proof is provided in Appendix A.
\end{proof}
\end{lemma}

\begin{lemma}
Define $\nabla \boldsymbol{F}_{m,r}=\sum_{i=1}^{I} \nabla F_m(\boldsymbol{\theta}_{m,r}^{i-1},\boldsymbol{\xi}_{m,r}^{i})$, and assume that the $j$-th entry of $ \nabla \boldsymbol{F}_{m,r}\in \mathbb{R}^d$ is within the range $[\underline{\nabla F}_{m,r,j},\overline{\nabla F}_{m,r,j}]$, the following two properties of stochastic quantization has been well established \cite{wang2021quantized,amiri2020federated}.
\begin{align*}
&\hspace{-2mm}\mathbb{E}_\mathcal{Q}\left[\mathcal{Q}(\Delta\boldsymbol{\theta}_{m,r}^{I})\right]=\Delta\boldsymbol{\theta}_{m,r}^{I} \numberthis \\
&\hspace{-2mm}\mathbb{E}_\mathcal{Q}\left[\lVert\mathcal{Q}(\Delta\boldsymbol{\theta}_{m,r}^{I})-\Delta\boldsymbol{\theta}_{m,r}^{I}\rVert^2\right]\leq \eta^2 J_{m,r}^2
\numberthis
\end{align*}
where $J_{m,r}^2\triangleq \frac{\delta_{m,r}^2}{(2^{B_{m,r}}-1)^2}$ and that $\delta_{m,r}\triangleq \sqrt{\frac{1}{4} \sum_{j=1}^{d}(\overline{\nabla F}_{m,r,j}-\underline{\nabla F}_{m,r,j})^2 }$. 
\end{lemma}

%
Based on the above assumptions and lemmas, we derive the following theorem which indicates that the optimality gap converges to 0 as $T\rightarrow \infty$.
\begin{theorem}
Let assumption 1$\sim$3 hold, choose $\eta=K^\star/(8LTI)^{\frac{1}{2}}$ and $I\leq (TI)^{\frac{1}{4}}/K^{\star\frac{3}{4}}$, by adopting our proposed method, it yields that
\begin{align*}   
&\frac{1}{T}\sum_{r=1}^T\mathbb{E}\left[\lVert\nabla F(\boldsymbol{\theta}_r) \rVert^2  \big\vert \mathcal{W}_r\neq \emptyset \right] \leq \\
&\frac{496 L}{11(TIK^{\star})^{\frac{1}{2}}}\left( \mathbb{E}\left[F(\boldsymbol{\theta}_0)\right]-F^\star \right)\\
&+\frac{31}{88(TI)^\frac{3}{2}K^{\star\frac{1}{2}}}\sum_{r=1}^{T}\sum_{m=1}^{M} \frac{1}{M} J_{m,r}^2\\
&+\left(\frac{39}{88(TIK^{\star})^{\frac{1}{2}}}+\frac{1}{88(TIK^{\star})^{\frac{3}{4}}}\right)\frac{\sigma^2}{b}+\\
&\Bigg( \frac{4}{11(TIK^{\star})^{\frac{1}{2}}}
+\frac{1}{22(TIK^{\star})^{\frac{3}{4}}} +\frac{31}{22(TI)^\frac{1}{4}K^{\star\frac{5}{4}}} \Bigg)\sum_{m=1}^{M}\frac{D_m^2}{M}.
\numberthis \label{eq:theorem 1}
\end{align*}
\begin{proof}
    The proof is provided in Appendix B.
\end{proof}
\end{theorem}

\section{Simulations}
We run experiments on the MNIST dataset, distribute an equal amount of data to each client, and fairly compare the performance of the following four methods. 
\begin{enumerate}[label=(\roman*)]
\item Our proposed method
 with $P_e$ under $\mathrm{SNR}=3$ per client and transmission rate $R=0.6$ respectively. 
\item Quantized FL (QFL) with perfect links\cite{amiri2020federated}, i.e., when $\mathrm{SNR}=\infty$. This benchmark provides insights into the ideal performance of an FL system. 
\item Anonymous (anon.) FL 
with the same $P_e$ in (i), where the PS is unaware of the identity of clients, such as amplify-and-forward. 
\item Non-anonymous (non-anon.) FL \cite{wang2021quantized} with same $P_e$ in (i). The updating rule of $\Delta{\boldsymbol{\theta}}_{r}$ follows (4) in \cite{wang2021quantized}.
\end{enumerate}

In the simulation, the number of clients is set to $M=10$. W.L.O.G, clients are equipped with the same stochastic quantizer with
$2^8-1$ quantization levels and boundary values of SQ are fixed. The number of training rounds is set to $T=20$, the number of local iterations $I=5$, the patch size per iteration is set to $b=1024$ and the learning rate is set to $\eta=0.01$. The classifier model is implemented using a 4-layer convolutional neural network (CNN) with SGD optimizer that consists of two convolution layers with 10 and 20 output channels respectively followed by 2 fully connected layers. 

\begin{figure}[!htb]
    \centering
    \begin{tikzpicture}[scale=1\columnwidth/10cm,font=\footnotesize,spy using outlines={circle, magnification=2,connect spies,dashed
}]
\begin{axis}[%
width=6cm,
height=3cm,
scale only axis,
xmin=1,
xmax=20,
ymin=10,
ymax=100,
xlabel={communication round},
ylabel={test accuracy$(\%)$},
y label style={at={(axis description cs:0.12,0.95)},
                     rotate=0, anchor=south east},
x label style={at={(axis description cs:0.7,-0.07)},
                     rotate=0, anchor=south east},                     
yminorticks=true,
xminorticks=true,
axis background/.style={fill=white},
xmajorgrids,
xmajorgrids,
ymajorgrids,
yminorgrids,
legend style={at={(-0.35,1.05)}, anchor=south west, legend columns=3, legend cell align=left, align=left, draw=white!15!black}
]

\addplot [color=cardinal, line width=1.5pt, mark size=1.5pt, mark=square*, mark options={solid, fill=cardinal, draw=cardinal}]
  table[row sep=crcr]{%
    1.0000    21.41\\
    2.0000    39.44\\
    3.0000    54.39\\
    4.0000    65.75\\
    5.0000    72.20\\
    6.0000    77.41\\
    7.0000    81.45\\
    8.0000    84.24\\
    9.0000    86.06\\
   10.0000    87.47\\
   11.0000    88.46\\
   12.0000    89.21\\
   13.0000    89.86\\
   14.0000    90.39\\
   15.0000    90.84\\
   16.0000    91.21\\
   17.0000    91.53\\
   18.0000    91.82\\
   19.0000    92.09\\
   20.0000    92.34\\
};
\addlegendentry{QFL (ideal)}

\addplot [color=dollarbill, line width=1pt, mark size=1pt, mark=diamond*, mark options={solid, fill=dollarbill, draw=dollarbill},dashed]
  table[row sep=crcr]{%
1	24.9840000000000\\
2	39.9110000000000\\
3	51.5480000000000\\
4	62.9360000000000\\
5	71.7540000000000\\
6	77.8260000000000\\
7	81.4810000000000\\
8	83.8190000000000\\
9	85.4110000000000\\
10	86.6160000000000\\
11	87.5640000000000\\
12	88.3140000000000\\
13	88.9400000000000\\
14	89.4880000000000\\
15	89.9630000000000\\
16	90.3930000000000\\
17	90.7520000000000\\
18	91.0680000000000\\
19	91.3730000000000\\
20	91.6000000000000\\
};
\addlegendentry{Non-anon. FL, SNR=5}

\addplot [color=dodgerblue, line width=1pt, mark size=1.5pt, mark=*, mark options={solid, fill=dodgerblue, draw=dodgerblue}]
  table[row sep=crcr]{%
    1.0000    23.5042105263158\\
    2.0000    40.1342105263158\\
    3.0000    55.2836842105263\\
    4.0000    65.5747368421053\\
    5.0000    72.8557894736842\\
    6.0000    78.1321052631579\\
    7.0000    81.7921052631579\\
    8.0000    84.1489473684211\\
    9.0000    85.8794736842106\\
   10.0000    87.1042105263158\\
   11.0000    88.0463157894737\\
   12.0000    88.7510526315790\\
   13.0000    89.3700000000000\\
   14.0000    89.8763157894737\\
   15.0000    90.3305263157895\\
   16.0000    90.7163157894737\\
   17.0000    91.0642105263158\\
   18.0000    91.3626315789474\\
   19.0000    91.6284210526316\\
   20.0000    91.8873684210527\\
};
\addlegendentry{Proposed, SNR=3}

\addplot [color=dollarbill, line width=1pt, mark size=1pt, mark=diamond*, mark options={solid, fill=dollarbill, draw=dollarbill}]
  table[row sep=crcr]{%
1	22.8700000000000\\
2	40.5810000000000\\
3	55.3350000000000\\
4	64.9460000000000\\
5	72.0370000000000\\
6	77.1450000000000\\
7	81.0280000000000\\
8	83.6730000000000\\
9	85.2810000000000\\
10	86.5470000000000\\
11	87.5360000000000\\
12	88.3550000000000\\
13	89.0220000000000\\
14	89.6060000000000\\
15	90.0480000000000\\
16	90.4500000000000\\
17	90.8410000000000\\
18	91.2220000000000\\
19	91.5420000000000\\
20	91.8040000000000\\
};
\addlegendentry{Non-anon. FL, SNR=3}

\addplot [color=dodgerblue, line width=1pt, mark size=1.5pt, mark=*, mark options={solid, fill=dodgerblue, draw=dodgerblue},dashed]
  table[row sep=crcr]{%
    1.0000    22.8023529411765\\
    2.0000    36.4188235294118\\
    3.0000    50.1388235294118\\
    4.0000    61.2370588235294\\
    5.0000    69.9111764705882\\
    6.0000    76.4647058823530\\
    7.0000    80.9582352941177\\
    8.0000    83.6023529411765\\
    9.0000    85.4235294117647\\
   10.0000    86.6852941176471\\
   11.0000    87.7247058823529\\
   12.0000    88.5111764705882\\
   13.0000    89.1711764705882\\
   14.0000    89.6970588235294\\
   15.0000    90.1941176470588\\
   16.0000    90.5770588235294\\
   17.0000    90.9517647058823\\
   18.0000    91.2747058823530\\
   19.0000    91.5594117647059\\
   20.0000    91.8341176470588\\
};
\addlegendentry{Proposed, SNR=5}

\addplot [color=burntorange, line width=1pt, mark size=1.5pt, mark=triangle*, mark options={solid, fill=burntorange, draw=burntorange},dashed]
  table[row sep=crcr]{%
    1.0000    17.83\\
    2.0000    29.06\\
    3.0000    36.45\\
    4.0000    43.68\\
    5.0000    50.88\\
    6.0000    56.72\\
    7.0000    61.11\\
    8.0000    67.39\\
    9.0000    70.54\\
   10.0000    74.07\\
   11.0000    77.16\\
   12.0000    79.17\\
   13.0000    81.49\\
   14.0000    82.71\\
   15.0000    83.72\\
   16.0000    84.38\\
   17.0000    84.89\\
   18.0000    85.13\\
   19.0000    85.50\\
   20.0000    85.72\\
};
\addlegendentry{Anon. FL, SNR=5}

\addplot [color=burntorange, line width=1pt, mark size=1.5pt, mark=triangle*, mark options={solid, fill=burntorange, draw=burntorange}]
  table[row sep=crcr]{%
1.0000    20.85\\
    2.0000    28.23\\
    3.0000    35.04\\
    4.0000    42.19\\
    5.0000    45.41\\
    6.0000    48.82\\
    7.0000    52.04\\
    8.0000    55.09\\
    9.0000    59.16\\
   10.0000    62.68\\
   11.0000    65.38\\
   12.0000    66.70\\
   13.0000    67.24\\
   14.0000    67.97\\
   15.0000    68.20\\
   16.0000    69.93\\
   17.0000    70.20\\
   18.0000    71.56\\
   19.0000    73.03\\
   20.0000    74.38\\
};
\addlegendentry{Anon. FL, SNR=3}

\coordinate (spypoint) at (axis cs:5,67);
\coordinate (spyviewer) at (axis cs:13,30);
\spy[width=1.5cm,height=1.5cm] on (spypoint) in node [fill=white] at (spyviewer);

\end{axis}
\end{tikzpicture} 
    \caption{Test accuracy comparison of four methods with $R=0.6$ under different SNRs in terms of communication round in the i.i.d. setting.} 
    \vspace{-6mm}
    \label{fig:test_acc_iid}
\end{figure}

\begin{figure}[!htb]
    \centering
    \begin{tikzpicture}[scale=1\columnwidth/10cm,font=\footnotesize,spy using outlines={circle, magnification=2,connect spies}]
\begin{axis}[%
width=6cm,
height=3cm,
scale only axis,
xmin=1,
xmax=20,
ymin=10,
ymax=90,
xlabel={communication round},
ylabel={test accuracy$(\%)$},
y label style={at={(axis description cs:0.12,0.95)},
                     rotate=0, anchor=south east},
x label style={at={(axis description cs:0.7,-0.07)},
                     rotate=0, anchor=south east},    
yminorticks=true,
xminorticks=true,
axis background/.style={fill=white},
xmajorgrids,
xmajorgrids,
ymajorgrids,
yminorgrids,
legend style={at={(-0.28,1.05)}, anchor=south west, legend columns=3, legend cell align=left, align=left, draw=white!15!black}
]

\addplot [color=cardinal, line width=1.5pt, mark size=1.5pt, mark=square*, mark options={solid, fill=cardinal, draw=cardinal}]
  table[row sep=crcr]{%
1	  17.5235714285714\\
2	  30.8921428571429\\
3	  41.2807142857143\\
4	  49.9778571428571\\
5	  57.4242857142857\\
6	  62.5050000000000\\
7	  66.2507142857143\\
8	  68.8500000000000\\
9	  70.7992857142857\\
10	  72.2892857142857\\
11	  73.6764285714286\\
12	  74.6621428571429\\
13	  75.6078571428572\\
14	  76.3764285714286\\
15	  77.0357142857143\\
16	  77.4921428571429\\
17	  78.1621428571429\\
18	  78.6207142857143\\
19	  79.0728571428571\\
20	  79.3964285714286\\
};
\addlegendentry{QFL (ideal), 5 cls.}

\addplot [color=cardinal, line width=1.5pt, mark size=1.5pt, mark=square*, mark options={solid, fill=cardinal, draw=cardinal},dashed]
  table[row sep=crcr]{%
1	11.9680000000000\\
2	17.9195000000000\\
3	24.9385000000000\\
4	30.5720000000000\\
5	34.6860000000000\\
6	38.7145000000000\\
7	42.4175000000000\\
8	45.9615000000000\\
9	48.5875000000000\\
10	51.7310000000000\\
11	54.4870000000000\\
12	56.9365000000000\\
13	58.9805000000000\\
14	60.5950000000000\\
15	62.2250000000000\\
16	63.5675000000000\\
17	64.7665000000000\\
18	65.7530000000000\\
19	67.0435000000000\\
20	67.5590000000000\\
};
\addlegendentry{QFL (ideal), 1 cl.}

\addplot [color=burntorange, line width=1pt, mark size=1.5pt, mark=triangle*, mark options={solid, fill=burntorange, draw=burntorange}]
  table[row sep=crcr]{%
1	14.6100000000000\\
2	11.2550000000000\\
3	10.5670000000000\\
4	11.4480000000000\\
5	11.1320000000000\\
6	10.6210000000000\\
7	10.6650000000000\\
8	10.8370000000000\\
9	10.4730000000000\\
10	10.8220000000000\\
11	10.3550000000000\\
12	10.1630000000000\\
13	10.4360000000000\\
14	10.6500000000000\\
15	10.7780000000000\\
16	10.6480000000000\\
17	10.5150000000000\\
18	10.4790000000000\\
19	10.7050000000000\\
20	10.8720000000000\\
};
\addlegendentry{Anon. FL, 5 cls.}

\addplot [color=burntorange, line width=1pt, mark size=1.5pt, mark=triangle*, mark options={solid, fill=burntorange, draw=burntorange},dashed]
  table[row sep=crcr]{%
1	10.7110000000000\\
2	12.7170000000000\\
3	10.4050000000000\\
4	10.4400000000000\\
5	11.3610000000000\\
6	10.2080000000000\\
7	10.1440000000000\\
8	10.3630000000000\\
9	11.1870000000000\\
10	10.3550000000000\\
11	10.4980000000000\\
12	10.3080000000000\\
13	10.1550000000000\\
14	10.3540000000000\\
15	10.6130000000000\\
16	10.4420000000000\\
17	10.3380000000000\\
18	10.2720000000000\\
19	10.7830000000000\\
20	10.5730000000000\\
};
\addlegendentry{Anon. FL, 1 cl.}

\addplot [color=dodgerblue, line width=1pt, mark size=1.5pt, mark=*, mark options={solid, fill=dodgerblue, draw=dodgerblue}]
  table[row sep=crcr]{%
1   17.4513157894737\\
2   30.8189473684210\\
3   41.0368421052631\\
4   49.2563157894737\\
5   55.8013157894737\\
6   61.0573684210526\\
7   64.8281578947369\\
8   67.5573684210526\\
9   69.7397368421053\\
10  71.4671052631579\\
11  72.9436842105263\\
12  74.1294736842105\\
13  75.2300000000000\\
14  76.1489473684211\\
15  76.9392105263158\\
16  77.6636842105263\\
17  78.2784210526316\\
18  78.7957894736842\\
19  79.3328947368421\\
20  79.8281578947368\\
};
\addlegendentry{Proposed, 5 cls.}

\addplot [color=dodgerblue, line width=1pt, mark size=1.5pt, mark=*, mark options={solid, fill=dodgerblue, draw=dodgerblue},dashed]
  table[row sep=crcr]{%
1	11.5331034482759\\
2	19.2375862068966\\
3	25.9727586206897\\
4	30.6603448275862\\
5	34.9951724137931\\
6	38.8072413793103\\
7	42.4527586206897\\
8	45.2351724137931\\
9	48.4337931034483\\
10	50.9693103448276\\
11	53.4696551724138\\
12	56.0100000000000\\
13	57.9627586206897\\
14	59.8993103448276\\
15	61.6151724137931\\
16	62.8772413793103\\
17	64.2158620689655\\
18	65.4172413793103\\
19	66.1800000000000\\
20	67.2551724137931\\
};
\addlegendentry{Proposed,1 cl.}

\addplot [color=dollarbill, line width=1pt, mark size=1pt, mark=diamond*, mark options={solid, fill=dollarbill, draw=dollarbill}]
  table[row sep=crcr]{%
1	15.2233333333333\\
2	25.4466666666667\\
3	34.9866666666667\\
4	41.8461111111111\\
5	48.5794444444445\\
6	54.6188888888889\\
7	56.8116666666667\\
8	61.7300000000000\\
9	62.7244444444445\\
10	61.5594444444444\\
11	65.4438888888889\\
12	69.1111111111111\\
13	70.2583333333333\\
14	69.2133333333333\\
15	70.8094444444445\\
16	72.9577777777778\\
17	72.0050000000000\\
18	71.3416666666667\\
19	73.5811111111111\\
20	74.4794444444445\\
};
\addlegendentry{Non-anon. FL, 5 cls.}

\addplot [color=dollarbill, line width=1pt, mark size=1pt, mark=diamond*, mark options={solid, fill=dollarbill, draw=dollarbill},dashed]
  table[row sep=crcr]{%
1	13.5425000000000\\
2	12.5183333333333\\
3	12.8441666666667\\
4	12.9366666666667\\
5	16.5408333333333\\
6	14.7733333333333\\
7	20.2325000000000\\
8	22.0925000000000\\
9	20.1475000000000\\
10	20.5841666666667\\
11	22.8075000000000\\
12	22.7375000000000\\
13	29.6858333333333\\
14	28.9200000000000\\
15	27.1600000000000\\
16	30.2075000000000\\
17	26.8383333333333\\
18	36.1141666666667\\
19	32.3950000000000\\
20	35.7000000000000\\
};
\addlegendentry{Non-anon. FL, 1 cl.}

\end{axis}
\end{tikzpicture} 
    \caption{Test accuracy comparison of four methods under different levels of data imbalances in terms of communication round with $\mathrm{SNR}=3$, $R=0.6$.} \vspace{-4mm}
    \label{fig:test_acc_noniid_new}
\end{figure}

The average test accuracy of the global model over multiple runs at each round is plotted in Fig. \ref{fig:test_acc_iid} and Fig. \ref{fig:test_acc_noniid_new} for the i.i.d settings and the non-i.i.d. settings, respectively. In the i.i.d setting, the training samples are shuffled and uniformly assigned to all clients, while in the non-i.i.d. setting, each client is allocated with $5$ classes (cls.) and $1$ class (cl.) of data respectively to achieve different levels of data imbalance. Our proposed method can achieve the optimal performance of an FL system with perfect connectivity, i.e., QFL, in both i.i.d. and non-i.i.d. settings. Although non-anon. FL can also achieve good performance in the i.i.d. setting due to homogeneous data distribution across clients, its performance (e.g., convergence speed and test accuracy) degrades significantly in the non-i.i.d. setting with the increased level of data imbalance, as the data distribution on the received clients does not fully reflect the overall data distribution on all clients. Compared with non-anon. FL, our proposed method improves test accuracy by 7\% and 21\% under different data imbalances respectively. The anon. FL works poorly in both i.i.d. and non-i.i.d. settings since the global model is severely distorted by communication outages. Compared with anon. FL and non-anon. FL, our proposed method shows strong robustness to frequent stragglers in all settings.

\section{Conclusion}
In this paper, we proposed a robust coded semi-decentralized network for FL, which does not necessitate any prior information and is suitable for practical scenarios. Our work is the first to exploit coded diversity to mitigate stragglers in FL over wireless networks and remains applicable to other scenarios, e.g., large language models (LLM) training.


\balance

\onecolumn
{\appendices
\section{Proof of Lemma 1}
\begin{proof}[Proof 1]
    W.L.O.G., assume that the probability of PS being unable to see local model update from client $m$ is $q$ through the i.i.d. symmetric network. Then, it can be written that
\begin{align*}
    &\mathbb{E}_{\mathcal{W}_r}\left[ \sum_{m\in \mathcal{W}_r} \frac{1}{\lvert \mathcal{W}_r\rvert}  \Delta\boldsymbol{\theta}_{m,r}^I\Bigg \vert \mathcal{W}_r\neq \emptyset  \right]\\
    &=\sum_{v=1}^{M} \sum_{\substack{\mathcal{W}_r\cap \bar{\mathcal{W}}_r=[M] \\ \lvert \mathcal{W}_r \rvert=v, \lvert  \bar{\mathcal{W}}_r \rvert=M-v }} \frac{(1-q)^vq^{M-v}}{1-q^M} \frac{1}{v} \sum_{m\in \mathcal{W}_r} \Delta\boldsymbol{\theta}_{m,r}^I\\
    &=\sum_{v=1}^{M}  \frac{(1-q)^vq^{M-v}}{1-q^M} \frac{1}{v} \sum_{\substack{\mathcal{W}_r\cap \bar{\mathcal{W}}_r=[M] \\ \lvert \mathcal{W}_r \rvert=v, \lvert  \bar{\mathcal{W}}_r \rvert=M-v }} \sum_{m\in \mathcal{W}_r} \Delta\boldsymbol{\theta}_{m,r}^I\\
    &\overset{(a)}=\sum_{v=1}^{M}  \frac{(1-q)^vq^{M-v}}{1-q^M} \frac{1}{v} \sum_{m=1}^M {M-1 \choose v-1}\Delta\boldsymbol{\theta}_{m,r}^I\\
    &\overset{(b)}=\sum_{v=1}^{M} {M \choose v} \frac{(1-q)^vq^{M-v}}{1-q^M} \frac{1}{M} \sum_{m=1}^M \Delta\boldsymbol{\theta}_{m,r}^I\\
    &=\frac{1}{M} \sum_{m=1}^M \Delta\boldsymbol{\theta}_{m,r}^I,
    \numberthis
    \label{eq:lemma1-a}
\end{align*}
where equality $(a)$ holds because each client is counted ${M-1 \choose v-1}$ times by summing over all possible sets $\mathcal{W}_r$, equality $(b)$ is due to $\sum_{v=1}^{M} {M \choose v} \frac{(1-q)^vq^{M-v}}{1-q^M}=1$, since that $\sum_{v=0}^{M} {M \choose v} (1-q)^vq^{M-v}=1$ and that ${M \choose 0} (1-q)^0q^{M-0}=q^M$. 
\end{proof}

\begin{proof}[Proof 2]
    Next, we prove the second conclusion in Lemma 1. Similar to the proof in (\ref{eq:lemma1-a}), it holds that
\begin{subequations}
\begin{align}
    &\mathbb{E}_{\mathcal{W}_r}\left[\sum_{m\in \mathcal{W}_r} \frac{1}{\lvert \mathcal{W}_r\rvert^2} \Delta\boldsymbol{\theta}_{m,r}^I\Bigg \vert \mathcal{W}_r\neq \emptyset\right] \label{eq:lemma1-b1}\\
    &=\sum_{v=1}^{M} \sum_{\substack{\mathcal{W}_r\cap \bar{\mathcal{W}}_r=[M] \\ \lvert \mathcal{W}_r \rvert=v, \lvert  \bar{\mathcal{W}}_r \rvert=M-v }} \frac{(1-q)^vq^{M-v}}{1-q^M} \frac{1}{v^2} \sum_{m\in \mathcal{W}_r} \Delta\boldsymbol{\theta}_{m,r}^I\label{eq:lemma1-b2}\\
    &=\sum_{v=1}^{M} \frac{1}{v} {M \choose v} \frac{(1-q)^vq^{M-v}}{1-q^M} \frac{1}{M} \sum_{m=1}^M \Delta\boldsymbol{\theta}_{m,r}^I \label{eq:lemma1-b3}\\
    &\triangleq\sum_{m=1}^M \Bar{\alpha}_m \boldsymbol{\theta}_{m,r}^I.
    \label{eq:lemma1-b4}
\end{align}
\end{subequations}
Let $\Delta\boldsymbol{\theta}_{m,r}^I=1$, comparison between (\ref{eq:lemma1-b1}) and (\ref{eq:lemma1-b3}) gives 
\begin{align*}
    \frac{1}{\bar{K}}\triangleq\mathbb{E}\left[ \frac{1}{\lvert \mathcal{W}_r\rvert} \Bigg \vert \mathcal{W}_r\neq \emptyset\right]=\sum_{v=1}^{M} \frac{1}{v} {M \choose v} \frac{(1-q)^vq^{M-v}}{1-q^M}.
    \numberthis
    \label{eq:bar_k}
\end{align*}
Substitute (\ref{eq:bar_k}) into (\ref{eq:lemma1-b3}), we obtain the relation between $\Bar{\alpha}_m$ and $\Bar{K}$.
\begin{align*}
    \Bar{\alpha}_m=\frac{1}{M\Bar{K}}
        \numberthis
    \label{eq:bar_a}
\end{align*}
\end{proof}

\begin{proof}[Proof 3]
Now, we prove the upper bound for the term $\frac{1}{\bar{K}}$.
\begin{align*}
    &\sum_{v=1}^{M} \frac{1}{v} {M \choose v} (1-q)^vq^{M-v}\\
    &=\sum_{v=1}^{M}\frac{1+v}{v}\frac{1}{1+v}{M \choose v}(1-q)^vq^{M-v}\\
    &\overset{(c)}=\sum_{v=1}^{M}\frac{1+v}{v}\frac{1}{(M+1)(1-q)}{M+1 \choose v+1}(1-q)^{v+1}q^{M-v}\\
    &\overset{(d)}\leq 2\sum_{v=1}^{M}\frac{1}{(M+1)(1-q)}{M+1 \choose v+1}(1-q)^{v+1}q^{M-v}\\
    &=\frac{2}{(M+1)(1-q)}\sum_{v=2}^{M+1}{M+1 \choose v}(1-q)^{v}q^{M-v+1}\\
    &\overset{(e)}=\frac{2}{(M+1)(1-q)}\left(1-q^{M+1}-(M+1)(1-q)q^M\right)\\
    &\leq\frac{2}{(M+1)(1-q)}
    \numberthis
    \label{eq:upper_K-1}
\end{align*}
where equality $(c)$ is due to $\frac{1}{1+v}{M \choose v}=\frac{1}{M+1}{M+1 \choose v+1}$, inequality $(d)$ is due to $\frac{1+v}{v}\leq 2$ since $v\geq 1$ in the sum, equality $(e)$ is due to $\sum_{v=0}^{M+1}{M+1 \choose v}(1-q)^{v}q^{M-v+1}=1$.

By (\ref{eq:upper_K-1}) and (\ref{eq:bar_k}), we can bound $\frac{1}{\bar{K}}$ approximately as
\begin{align*}
    \frac{1}{\bar{K}}&\leq \frac{2}{(M+1)(1-q)(1-q^M)}\\
    &\overset{(f)}\approx \frac{2}{(M+1)(1-P_e^{2M-1})(1-P_e^{M(2M-1)})}\triangleq \frac{1}{K^{\star}},
    \numberthis
    \label{eq:upper_K-2}
\end{align*}
where inequality $(f)$ is simply due to the fact $q\approx P_e^{2M-1}\ll 1$.
\end{proof}

\section{Proof of Theorem 1}
The proof of Theorem 1 closely follows the proof of Theorem 1 in \cite{wang2021quantized}, which bounds the optimality gap of QFL in the presence of communication stragglers with client sampling. Here, we will only provide the sketch proof and detail the differing aspects.

By A.1, we have 
\begin{align*}
      &\mathbb{E}\left[F(\boldsymbol{\theta}_{r+1})\right]-\mathbb{E}\left[F(\boldsymbol{\theta}_{r})\right] 
     \overset{\text{A.1}}\leq\mathbb{E}\left[\left\langle \nabla F(\boldsymbol{\theta}_{r}), \boldsymbol{\theta}_{r+1}-\boldsymbol{\theta}_{r} \right\rangle\right]
      +\frac{L}{2}\mathbb{E}\left[\lVert \boldsymbol{\theta}_{r+1}-\boldsymbol{\theta}_{r} \rVert^2\right]. \label{eq:theorem 1_uniform_tau}
      \numberthis
\end{align*}
The following three key lemmas are required for the complete proof of Theorem 1.
\begin{lemma}
Under A.1$\sim$A.3, it holds that 
\begin{align*}
    &\mathbb{E}\left[\left\langle \nabla F(\boldsymbol{\theta}_{r}), \sum_{m\in \mathcal{W}_r} \frac{1}{\lvert \mathcal{W}_r\rvert} \mathcal{Q} (\Delta\boldsymbol{\theta}_{m,r}^I) \right\rangle\right]\\
    &\leq  -\frac{\eta I}{2}\mathbb{E}\left[ \left\lVert \nabla F(\boldsymbol{\theta}_{r}) \right\rVert^2 \right]
     +\frac{\eta L^2}{M}\sum_{m=1}^{M} \sum_{i=1}^{I} \mathbb{E}\left[ \left\lVert \boldsymbol{\theta}_{r}- \boldsymbol{\theta}_{m,r}^{i-1} \right\rVert^2 \right].
     \numberthis
\end{align*}    
\begin{proof}
    The proof is provided in Appendix \ref{Proof of Lemma 3}. 
\end{proof}
\end{lemma}
\begin{lemma}
Under A.1$\sim$A.3, it holds that 
\begin{align*}
    \mathbb{E}\left[ \left\lVert \boldsymbol{\theta}_{r+1}-\boldsymbol{\theta}_{r} \right\rVert^2 \right]
    &\leq \frac{I\sigma^2}{b\bar{K}}+\eta^2 \sum_{m=1}^{M} \bar{\alpha}_m J_{m,r}^2+\frac{2IL^2}{M}  \mathbb{E}\left[\left\lVert\boldsymbol{\theta}_{m,r}^{i-1}-\boldsymbol{\theta}_{r} \right\rVert^2 \right]\\
    &+ 4I^2\sum_{m=1}^{M}\bar{\alpha}_m D_m^2+4I^2\mathbb{E}\left[ \left\lVert\nabla F(\boldsymbol{\theta}_{r}) \right\rVert^2 \right]
    \numberthis
\end{align*}

\begin{proof}
    The proof is provided in Appendix \ref{Proof of Lemma 4}. 
\end{proof}
\end{lemma}
\begin{lemma}
Under A.1$\sim$A.3, it holds that 
\begin{align*}
    &\sum_{i=1}^{I} \mathbb{E}\left[ \left\lVert \boldsymbol{\theta}_{m,r+1}^{i-1}-\boldsymbol{\theta}_r\right\rVert^2 \right]
    \leq \frac{\eta^2 I^3 \frac{\sigma^2}{b} + 4\eta^2E^3D_m^2 + 4\eta^2 E^3 \mathbb{E}\left[ \left\lVert \nabla F(\boldsymbol{\theta}_{r}) \right\rVert^2 \right] }{1-2\eta^2 I^2L^2}     
    \numberthis
\end{align*}
\begin{proof}
    The proof can be found in Appendix B of \cite{wang2021quantized}. 
\end{proof}
\end{lemma}

Our Lemma 3, Lemma 4, and Lemma 5 are mathematically the same as Lemma 3$\sim$5 in \cite{wang2021quantized} when $q$ is identical and $p_i=\frac{1}{M}$ in \cite{wang2021quantized}. Thus, our theorem can be derived the same way as the corollary 1 in \cite{wang2021quantized}. That is, 
Let assumption 1$\sim$3 hold, choose $\eta=K/(8LTI)^{\frac{1}{2}}$ and $I\leq (TI)^{\frac{1}{4}}/K^{\frac{3}{4}}$, by adopting our proposed method, it yields that
\begin{align*}   
&\frac{1}{T}\sum_{r=1}^T\mathbb{E}\left[\lVert\nabla F(\boldsymbol{\theta}_r) \rVert^2  \big\vert \mathcal{W}_r\neq \emptyset \right] \leq 
\frac{496 L}{11(TIK)^{\frac{1}{2}}}\left( \mathbb{E}\left[F(\boldsymbol{\theta}_0)\right]-F^\star \right)\\
&+\frac{31}{88(TI)^\frac{3}{2}K^{\frac{1}{2}}}\sum_{r=1}^{T}\sum_{m=1}^{M} \frac{1}{M} J_{m,r}^2
+\left(\frac{39}{88(TIK)^{\frac{1}{2}}}+\frac{1}{88(TIK)^{\frac{3}{4}}}\right)\frac{\sigma^2}{b}\\
&+\left( \frac{4}{11(TIK)^{\frac{1}{2}}}+\frac{1}{22(TIK)^{\frac{3}{4}}} +\frac{31}{22(TI)^\frac{1}{4}K^{\frac{5}{4}}} \right)\sum_{m=1}^{M}\frac{1}{M} D_m^2.
\numberthis \label{eq:pre-theorem}
\end{align*}
Since $\frac{1}{\bar{K}}\leq \frac{1}{K^\star}$, (\ref{eq:pre-theorem}) can be further upper bounded by 
\begin{align*}
&\frac{1}{T}\sum_{r=1}^T\mathbb{E}\left[\lVert\nabla F(\boldsymbol{\theta}_r) \rVert^2  \big\vert \mathcal{W}_r\neq \emptyset \right] \leq 
\frac{496 L}{11(TIK^{\star})^{\frac{1}{2}}}\left( \mathbb{E}\left[F(\boldsymbol{\theta}_0)\right]-F^\star \right)\\
&+\frac{31}{88(TI)^\frac{3}{2}K^{\star\frac{1}{2}}}\sum_{r=1}^{T}\sum_{m=1}^{M} \frac{1}{M} J_{m,r}^2
+\left(\frac{39}{88(TIK^{\star})^{\frac{1}{2}}}+\frac{1}{88(TIK^{\star})^{\frac{3}{4}}}\right)\frac{\sigma^2}{b}\\
&+\left( \frac{4}{11(TIK^{\star})^{\frac{1}{2}}}+\frac{1}{22(TIK^{\star})^{\frac{3}{4}}} +\frac{31}{22(TI)^\frac{1}{4}K^{\star\frac{5}{4}}} \right)\sum_{m=1}^{M}\frac{1}{M} D_m^2,
\numberthis \label{eq:pre-theorem2}
\end{align*}
where the right side of (\ref{eq:pre-theorem2}) corresponds to 
 $\eta=K^{\star}/(8LTI)^{\frac{1}{2}}$ and $I\leq (TI)^{\frac{1}{4}}/K^{\star\frac{3}{4}}$.

\section{Proof of Lemma 3}\label{Proof of Lemma 3}
\begin{proof}
    \begin{align*}
     &\mathbb{E}\left[\left\langle \nabla F(\boldsymbol{\theta}_{r}), \sum_{m\in \mathcal{W}_r} \frac{1}{\lvert \mathcal{W}_r\rvert} \mathcal{Q} (\Delta\boldsymbol{\theta}_{m,r}^I) \right\rangle\right]\\
     &\overset{(g)}=\mathbb{E}\left[\left\langle \nabla F(\boldsymbol{\theta}_{r}), \frac{1}{M}\sum_{m=1}^{M} \mathcal{Q} (\Delta\boldsymbol{\theta}_{m,r}^I) \right\rangle\right]\\
     &\overset{(h)}=\mathbb{E}\left[\left\langle \nabla F(\boldsymbol{\theta}_{r}), \frac{1}{M}\sum_{m=1}^{M} \Delta\boldsymbol{\theta}_{m,r}^I \right\rangle\right]\\
     &\overset{\mathrm{A.2}}=-\eta\sum_{i=1}^{I}\mathbb{E}\left[\left\langle \nabla F(\boldsymbol{\theta}_{r}), \frac{1}{M}\sum_{m=1}^{M} \nabla F_m(\boldsymbol{\theta}_{m,r}^{i-1}) \right\rangle\right]\\
     &\overset{(i)}=-\frac{\eta}{2}\sum_{i=1}^{I}\mathbb{E}\left[ \left\lVert \nabla F(\boldsymbol{\theta}_{r}) \right\rVert^2 \right]
     -\frac{\eta}{2}\sum_{i=1}^{I}\mathbb{E}\left[ \left\lVert \frac{1}{M}\sum_{m=1}^{M} \nabla F_m(\boldsymbol{\theta}_{m,r}^{i-1}) \right\rVert^2 \right]\\
     &\hspace{4mm}+\frac{\eta}{2}\sum_{i=1}^{I} \mathbb{E}\left[ \left\lVert \nabla F(\boldsymbol{\theta}_{r})-\frac{1}{M}\sum_{m=1}^{M} \nabla F_m(\boldsymbol{\theta}_{m,r}^{i-1}) \right\rVert^2 \right]\\
     &\leq -\frac{\eta I}{2}\mathbb{E}\left[ \left\lVert \nabla F(\boldsymbol{\theta}_{r}) \right\rVert^2 \right]
     +\frac{\eta}{2}\sum_{i=1}^{I} \mathbb{E}\left[ \left\lVert \nabla F(\boldsymbol{\theta}_{r})-\frac{1}{M}\sum_{m=1}^{M} \nabla F_m(\boldsymbol{\theta}_{m,r}^{i-1}) \right\rVert^2 \right]\\
     &\overset{(j)}\leq -\frac{\eta I}{2}\mathbb{E}\left[ \left\lVert \nabla F(\boldsymbol{\theta}_{r}) \right\rVert^2 \right]
     +\eta\sum_{i=1}^{I} \mathbb{E}\left[ \left\lVert \nabla F(\boldsymbol{\theta}_{r})-\frac{1}{M}\sum_{m=1}^{M} \nabla F_m(\boldsymbol{\theta}_{r}) \right\rVert^2 \right]\\
     &\hspace{4mm}+\eta\sum_{i=1}^{I} \mathbb{E}\left[ \left\lVert \frac{1}{M}\sum_{m=1}^{M}\nabla F_m(\boldsymbol{\theta}_{r})- \frac{1}{M}\sum_{m=1}^{M}\nabla F_m(\boldsymbol{\theta}_{m,r}^{i-1}) \right\rVert^2 \right]\\
     &\overset{(k)}= -\frac{\eta I}{2}\mathbb{E}\left[ \left\lVert \nabla F(\boldsymbol{\theta}_{r}) \right\rVert^2 \right]
     +\frac{\eta}{M}\sum_{m=1}^{M} \sum_{i=1}^{I} \mathbb{E}\left[ \left\lVert \nabla F_m(\boldsymbol{\theta}_{r})- \nabla F_m(\boldsymbol{\theta}_{m,r}^{i-1}) \right\rVert^2 \right]\\
     &\overset{\mathrm{A.1}}\leq  -\frac{\eta I}{2}\mathbb{E}\left[ \left\lVert \nabla F(\boldsymbol{\theta}_{r}) \right\rVert^2 \right]
     +\frac{\eta L^2}{M}\sum_{m=1}^{M} \sum_{i=1}^{I} \mathbb{E}\left[ \left\lVert \boldsymbol{\theta}_{r}- \boldsymbol{\theta}_{m,r}^{i-1} \right\rVert^2 \right],
     \numberthis
    \end{align*}
    where $(g)$ follows Lemma 1, $(h)$ follows Lemma 2, $(i)$ is due to the basic property $\langle a,b \rangle=\frac{1}{2}\lVert a \rVert^2+\frac{1}{2}\lVert b \rVert^2-\frac{1}{2}\lVert a-b \rVert^2$, $(j)$ is due to $ \lVert a+b \rVert^2 \leq 2\lVert a \rVert^2 +2\lVert b \rVert^2 $, $(j)$ is due to unbiased estimation and the fact that $\nabla F(\boldsymbol{\theta}_{r})=\frac{1}{M}\sum_{m=1}^{M} \nabla F_m(\boldsymbol{\theta}_{r})$. 
\end{proof}

\section{Proof of Lemma 4}\label{Proof of Lemma 4}
\begin{proof}
\begin{align*}
    &\mathbb{E}\left[ \left\lVert \boldsymbol{\theta}_{r+1}-\boldsymbol{\theta}_{r} \right\rVert^2 \right]\\
    &=\mathbb{E}\left[ \left\lVert \sum_{m\in \mathcal{W}_r}\frac{1}{\lvert \mathcal{W}_r \rvert} \mathcal{Q}(\Delta \boldsymbol{\theta}_{m,r}^I) \right\rVert^2 \right]\\
    &\overset{(l)}= \mathbb{E}\left[ \left\lVert \sum_{m\in \mathcal{W}_r}\frac{1}{\lvert \mathcal{W}_r \rvert} \Delta \boldsymbol{\theta}_{m,r}^I \right\rVert^2 \right]
    +\mathbb{E}\left[ \left\lVert \sum_{m\in \mathcal{W}_r}\frac{1}{\lvert \mathcal{W}_r \rvert} \left(\Delta \boldsymbol{\theta}_{m,r}^I-\mathcal{Q}(\Delta \boldsymbol{\theta}_{m,r}^I)\right) \right\rVert^2 \right]\\
    &\overset{(m)}=\eta^2 \underbrace{\mathbb{E}\left[ \left\lVert\sum_{m\in \mathcal{W}_r}\frac{1}{\lvert \mathcal{W}_r \rvert}\sum_{i=1}^{I}\left(\nabla F_m(\boldsymbol{\theta}_{m,r}^{i-1},\xi_{m,r}^i)-\nabla F_m(\boldsymbol{\theta}_{m,r}^{i-1})\right) \right\rVert^2 \right]}_{T_1}\\
    &\hspace{4mm}+\eta^2 \underbrace{\mathbb{E}\left[ \left\lVert\sum_{m\in \mathcal{W}_r}\frac{1}{\lvert \mathcal{W}_r \rvert}\sum_{i=1}^{I}\nabla F_m(\boldsymbol{\theta}_{m,r}^{i-1}) \right\rVert^2 \right]}_{T_2}
    +\underbrace{\mathbb{E}\left[ \left\lVert\sum_{m\in \mathcal{W}_r}\frac{1}{\lvert \mathcal{W}_r \rvert}\left( \mathcal{Q}(\Delta\boldsymbol{\theta}_{m,r}^{I})-\Delta\boldsymbol{\theta}_{m,r}^{I} \right) \right\rVert^2 \right]}_{T_3},
    \numberthis
    \label{eq:lemm4-1}
\end{align*}
where $(m)$ and $(n)$ both follows the fact that $\mathbb{E}[\lVert x \rVert^2]=\lVert \mathbb{E}[x]\rVert^2+\mathbb{E}[\lVert x-\mathbb{E}[x] \rVert^2]$. Additionally, $(n)$ also utilize Assumption 2.
Next, we provide the upper bounds for the terms $T_1$, and $T_3$ respectively.
\begin{align*}
    &T_1\overset{(n)}=\mathbb{E}\left[\frac{\sum_{m\in \mathcal{W}_r}\sum_{i=1}^{I}\left\lVert\left(\nabla F_m(\boldsymbol{\theta}_{m,r}^{i-1},\xi_{m,r}^i)-\nabla F_m(\boldsymbol{\theta}_{m,r}^{i-1})\right)\right\rVert^2}{\lvert \mathcal{W}_r\rvert^2 }  \right]\\
    &\hspace{4mm}\overset{A.2}\leq \mathbb{E}\left[ \frac{I\frac{\sigma^2}{b}}{\lvert \mathcal{W}_r\rvert} \right]=\frac{I\sigma^2}{b}\mathbb{E}\left[ \frac{1}{\lvert \mathcal{W}_r\rvert}\right]\overset{L.1}=\frac{I\sigma^2}{b\bar{K}},
    \numberthis
    \label{eq:bound T1}
\end{align*}
where $(n)$ is due to unbiased estimation.
\begin{align*}
    &T_3=\mathbb{E}\left[\frac{\sum_{m\in \mathcal{W}_r}\left\lVert  \mathcal{Q}(\Delta\boldsymbol{\theta}_{m,r}^{I})-\Delta\boldsymbol{\theta}_{m,r}^{I} \right\rVert^2}{\lvert \mathcal{W}_r\rvert^2 }  \right]
    \overset{L.2}\leq \mathbb{E}\left[\frac{\sum_{m\in \mathcal{W}_r} \eta^2 J_{m,r}^2 }{\lvert \mathcal{W}_r\rvert^2 }  \right]\overset{L.1}=\eta^2 \sum_{m=1}^{M} \bar{\alpha}_m J_{m,r}^2
    \numberthis
    \label{eq:bound T3}
\end{align*}
Now we focus on bounding $T_2$.
\begin{align*}
    T_2&=\mathbb{E}\left[ \left\lVert\sum_{m\in \mathcal{W}_r}\frac{1}{\lvert \mathcal{W}_r \rvert}\sum_{i=1}^{I}\nabla F_m(\boldsymbol{\theta}_{m,r}^{i-1}) \right\rVert^2 \right]\\
    &\overset{(o)}\leq 2\mathbb{E}\left[ \left\lVert\sum_{m\in \mathcal{W}_r}\frac{1}{\lvert \mathcal{W}_r \rvert}\sum_{i=1}^{I}\left(\nabla F_m(\boldsymbol{\theta}_{m,r}^{i-1})-\nabla F_m(\boldsymbol{\theta}_{r})\right) \right\rVert^2 \right]
    +2\mathbb{E}\left[ \left\lVert\sum_{m\in \mathcal{W}_r}\frac{1}{\lvert \mathcal{W}_r \rvert}\sum_{i=1}^{I}\nabla F_m(\boldsymbol{\theta}_{r}) \right\rVert^2 \right]\\
    &\overset{(p)}\leq 2I \mathbb{E}\left[\sum_{m\in \mathcal{W}_r}\frac{1}{\lvert \mathcal{W}_r \rvert}\sum_{i=1}^{I}\left\lVert\nabla F_m(\boldsymbol{\theta}_{m,r}^{i-1})-\nabla F_m(\boldsymbol{\theta}_{r}) \right\rVert^2 \right]
    +2\mathbb{E}\left[ \left\lVert\sum_{m\in \mathcal{W}_r}\frac{1}{\lvert \mathcal{W}_r \rvert}\sum_{i=1}^{I}\nabla F_m(\boldsymbol{\theta}_{r}) \right\rVert^2 \right]\\
    &\overset{L.1}\leq \frac{2I}{M}\sum_{m=1}^M\sum_{i=1}^{I} \mathbb{E}\left[\left\lVert\nabla F_m(\boldsymbol{\theta}_{m,r}^{i-1})-\nabla F_m(\boldsymbol{\theta}_{r}) \right\rVert^2 \right]
    +2\mathbb{E}\left[ \left\lVert\sum_{m\in \mathcal{W}_r}\frac{1}{\lvert \mathcal{W}_r \rvert}\sum_{i=1}^{I}\nabla F_m(\boldsymbol{\theta}_{r}) \right\rVert^2 \right]\\
    &\overset{A.1}\leq \frac{2IL^2}{M}  \mathbb{E}\left[\left\lVert\boldsymbol{\theta}_{m,r}^{i-1}-\boldsymbol{\theta}_{r} \right\rVert^2 \right]+\underbrace{2I^2\mathbb{E}\left[ \left\lVert\sum_{m\in \mathcal{W}_r}\frac{1}{\lvert \mathcal{W}_r \rvert}\nabla F_m(\boldsymbol{\theta}_{r}) \right\rVert^2 \right]}_{T_{2-1}},
    \numberthis
    \label{eq:bound T2}
\end{align*}
where $(o)$ follows basic property $ \lVert a+b \rVert^2 \leq 2\lVert a \rVert^2 +2\lVert b \rVert^2 $, and $(p)$ follows the convexity of $l_2$-norm.

Furthermore, we have
\begin{align*}
    T_{2-1}&\overset{(q)}=4I^2\mathbb{E}\left[ \left\lVert\sum_{m\in \mathcal{W}_r}\frac{1}{\lvert \mathcal{W}_r \rvert}\left(\nabla F_m(\boldsymbol{\theta}_{r})-F(\boldsymbol{\theta}_{r}) \right)\right\rVert^2 \right]
    +4I^2\mathbb{E}\left[ \left\lVert\nabla F(\boldsymbol{\theta}_{r}) \right\rVert^2 \right]\\
    &\overset{(r)}=4I^2\mathbb{E}\left[\frac{ \sum_{m\in \mathcal{W}_r} \left\lVert \nabla F_m(\boldsymbol{\theta}_{r})-F(\boldsymbol{\theta}_{r}) \right\rVert^2}{\lvert \mathcal{W}_r \rvert^2}\right]
    +4I^2\mathbb{E}\left[ \left\lVert\nabla F(\boldsymbol{\theta}_{r}) \right\rVert^2 \right]\\
    &\overset{A.3}\leq 4I^2\sum_{m=1}^{M}\bar{\alpha}_m D_m^2+4I^2\mathbb{E}\left[ \left\lVert\nabla F(\boldsymbol{\theta}_{r}) \right\rVert^2 \right],
    \numberthis
    \label{eq:bound T21}
\end{align*}
where $(q)$ follows $\mathbb{E}[\lVert x \rVert^2]=\lVert \mathbb{E}[x]\rVert^2+\mathbb{E}[\lVert x-\mathbb{E}[x] \rVert^2]$ and Lemma 1, $(r)$ is due to unbiased estimation given by Lemma 1 and Lemma 2 in \cite{wang2020tackling}. 
By substituting (\ref{eq:bound T21}) into (\ref{eq:bound T2}), we obtain the bound for $T_2$ as follows.
\begin{align*}
    T_2\leq &\frac{2IL^2}{M}  \mathbb{E}\left[\left\lVert\boldsymbol{\theta}_{m,r}^{i-1}-\boldsymbol{\theta}_{r} \right\rVert^2 \right]+ 4I^2\sum_{m=1}^{M}\bar{\alpha}_m D_m^2+4I^2\mathbb{E}\left[ \left\lVert\nabla F(\boldsymbol{\theta}_{r}) \right\rVert^2 \right]
    \numberthis
    \label{eq:bound T2}
\end{align*}
Substitute (\ref{eq:bound T1}), (\ref{eq:bound T2}), (\ref{eq:bound T3}) into (\ref{eq:lemm4-1}), we acquire lemma 4.

\end{proof}
}

\end{document}